\documentclass[12pt]{article}
\usepackage{amsmath}
\usepackage{amssymb}
\usepackage{a4}
\usepackage{epsfig}
\def\ltap{\raisebox{-.4ex}{\rlap{$\,\sim\,$}} \raisebox{.4ex}{$\,<\,$}} 
 
\newcommand\as{\alpha_{\mathrm{S}}} 
\newcommand{\mr}[1] {\mathrm{#1}}
\newcommand{\kfac} {$K$-factor}
\newcommand{\pt} {\ifmmode p_\mathrm{T}\else   $p_\mathrm{T}$\fi}
\newcommand{\ptH} {\ifmmode p_\mathrm{T}^\mathrm{H}\else   $p_\mathrm{T}^\mathrm{H}$\fi}
\newcommand{\ptWW} {\ifmmode p_\mathrm{T}^\mathrm{WW}\else   $p_\mathrm{T}^\mathrm{WW}$\fi}
\newcommand{\ptJ} {\ifmmode p_\mathrm{T}^\mathrm{jet}\else   $p_\mathrm{T}^\mathrm{jet}$\fi}
\newcommand{\ptmin} {\ifmmode p^\ell_\mathrm{T min}\else   $p^\ell_\mathrm{T min}$\fi}
\newcommand{\ptmax} {\ifmmode p^\ell_\mathrm{T max}\else   $p^\ell_\mathrm{T max}$\fi}

\newcommand{\newc}[1] {\newcommand{#1}}
\newc{\R}{$R$}
\newc{\charginom}{M_{\tilde \chi}^{+}}
\newc{\mue}{\mu_{\tilde{e}_{iL}}}
\newc{\mud}{\mu_{\tilde{d}_{jL}}}
\newc{\barr}{\begin{eqnarray}}
\newc{\earr}{\end{eqnarray}}
\newc{\beq}{\begin{equation}}
\newc{\eeq}{\end{equation}}
\newc{\ra}{\rightarrow}
\newc{\lam}{\lambda}
\newc{\eps}{\epsilon}
\newc{\gev}{\,GeV}
\newc{\tev}{\,TeV}
\newc{\eq}[1]{(\ref{eq:#1})}
\newc{\eqs}[2]{(\ref{eq:#1},\ref{eq:#2})}
\newc{\etal}{{\it et al.}\ }
\newc{\eg}{{\it e.g.}\ }
\newc{\ie}{{\it i.e.}\ }
\newc{\nonum}{\nonumber}
\newc{\lab}[1]{\label{eq:#1}}
\newc{\dpr}[2]{({#1}\cdot{#2})}
\newc{\gsim}{\stackrel{>}{\sim}}
\newc{\lsim}{\stackrel{<}{\sim}}

\newc{\gevcc}{GeV}
\newc{\tevcc}{TeV}
\newc{\gevc}{GeV}
\newc{\PY}{\textsc{PYTHIA}}

\begin{document}

\begin{titlepage}
\begin{flushright}
ETHZ-IPP  PR-04-01\\
CERN-PH-TH/2004-035\\
{April 21, 2004}
\end{flushright}
\vskip 2cm
\begin{center}
{\bf \Large
Effective \textit{K}-factors \\
\vskip .2cm
for \boldmath {$\mathbf { gg \rightarrow H \rightarrow WW \rightarrow  \ell \nu \ell \nu } $} at the LHC}
\end{center}
\smallskip \smallskip \bigskip
\begin{center}
{\large G. Davatz$^1$, G. Dissertori$^1$, M. Dittmar$^1$, M. Grazzini$^2$, F. Pauss$^1$}
\end{center}
\bigskip

\begin{center}
$^1$Institute for Particle Physics (IPP), ETH Zurich, \\
CH-8093 Zurich, Switzerland
\end{center}

\begin{center}
$^2$ Department of Physics, CERN\\
Theory Division\\
CH--1211 Geneva 23, Switzerland
\end{center}


\begin{abstract}
\noindent 
A simulation of the search for
the Standard Model Higgs boson at the LHC,
in the channel
$\mathrm{gg}\to \mathrm{H} \to \mathrm{WW} \to \ell \nu \ell \nu$,
is described. 
Higher-order QCD corrections are taken into account by
using a reweighting procedure, which 
allows us to combine
event rates obtained with 
the PYTHIA Monte Carlo program with  
the most up-to-date theoretical predictions for
the transverse-momentum spectra of the Higgs signal and its corresponding WW background.
With this method the discovery potential for Higgs masses between $140$ and $180$ GeV 
is recalculated and the potential statistical
significance of this channel is found to increase considerably. 
For a Higgs mass of 165 GeV a signal-to-background ratio 
of almost 2:1 can be obtained.
A statistical significance of five standard deviations might already be achieved 
with an integrated luminosity close to 0.4 fb$^{-1}$.
Using this approach,  an experimental effective \kfac\
 of about 2.04 is obtained for the considered Higgs signature, which is   
only about 15\% smaller than the theoretical inclusive \kfac.

\end{abstract}

\end{titlepage}

\renewcommand{\thefootnote}{\fnsymbol{footnote}}


\section{Introduction}

The design of the ATLAS and CMS experiments
at the LHC was guided by the requirement to  
have high sensitivity for discovering the Higgs boson
within the full mass range between 100 GeV and
approximately 1 TeV \cite{atlascms}.

In recent years a large effort has gone into accurate 
calculations of many Higgs signal and background cross sections,
which in most cases are now known to next-to-leading order (NLO)
accuracy \cite{leshouches}. 
For the dominant
Standard Model (SM)
Higgs production mechanism,
gluon--gluon fusion,
and as far as the total cross section is concerned, 
even next-to-next-to-leading order (NNLO) \cite{higgsnnlo} QCD corrections
have been computed\footnote{More precisely, the NNLO calculation has been performed
in the large $m_{\mathrm{top}}$ approximation.}.

By contrast,
and despite the recent considerable progress in
improving Monte Carlo (MC) event generators,
a complete MC program where the same higher-order
QCD corrections are included does not exist yet.

QCD corrections to signal and background cross sections are thus either
ignored or taken into account in a very naive approach. 
Normally, results obtained
with a standard MC program are simply scaled
with the so-called inclusive \kfac.
Although the \kfac\ should not be considered as a physical quantity,
this approach is believed to provide a reasonable
simulation environment, which allows us to study the acceptance for 
many signatures. In the context of Higgs searches,
an example is the decay of Higgs bosons
into four leptons, $\mr{H} \ra \mr{ZZ} \ra 4\ell$.
This signature 
is not really sensitive to additional jet activities,
and it is usually assumed that  
the search sensitivity 
depends mainly on signal and background cross sections,
the particular detector model
and the used selection criteria.
Consequently, a simple 
scaling of signal and background with the inclusive \kfac, according to the most accurate 
theoretical prediction, should give reasonable results. This assumption has recently been confirmed 
in a more quantitative study for the decay $\mr{H} \ra \mr{ZZ} \ra 4\ell$ \cite{cranmer03}.

However, the situation is different if,
in addition to some particle identification,
the event kinematics has to be exploited  
to separate the signal from backgrounds.
A typical example is the proposed Higgs search in the mass range $155$--$180$ \gevcc\ \cite{herbidittmar},
where the identification of the decay 
$\mr{H}\ra\mr{W}^{+}\mr{W}^{-} \ra \ell^{+}\nu \ell^{'-}{\bar\nu}$ 
requires a jet veto in order to remove $\mr{t}\bar{\mr{t}}$ events. 
In addition, other cuts are required, which 
exploit the spin correlations between the W bosons 
and the resulting transverse momentum (\pt) spectra of the charged leptons. 
These cuts are particularly sensitive
when the
Higgs mass is close to 2$M_\mr{W}$  
and if the Higgs boson is produced with small transverse momentum \ptH\ . 
Consequently,
we cannot expect
that the inclusive \kfac s can be used
directly and a more careful investigation is needed. 

The effects of a jet veto on the \kfac\ have been studied
in QCD perturbation theory up to NNLO
in Ref.\ \cite{Catani:2001cr}.
These results show that the impact of higher-order
QCD corrections is reduced if a jet veto is applied.
This study 
demonstrated clearly that the simple \kfac\ scaling
cannot be used in this case.
It is therefore
important
to determine the effective ``experimental'' \kfac\ 
in combination with a detailed simulation of cuts, for both signal and background.

In this paper we reconsider the
${\rm gg}\to\mr{H}\ra\mr{W}^{+}\mr{W}^{-} \ra \ell^{+}\nu \ell^{'-}{\bar\nu}$
channel by using a reweighting procedure
of events generated with the PYTHIA \cite{pythia} MC program.
The reweighting is performed according to the most up-to-date
theoretical predictions for \pt\ spectra of the Higgs \cite{Bozzi:2003jy}
and the non-resonant WW production, which is the most important background.
This method allows us to include higher-order QCD corrections 
in combination with experimental selection criteria to a good approximation.
  
The paper is organized as follows.
In Section \ref{defKfacs} the reweighting technique is introduced.
In Section \ref{selection} we use PYTHIA to determine the detection efficiency  
as a function of \ptH, using selection criteria 
based on the ideas given in \cite{herbidittmar}. These selection criteria are 
relatively robust and it can be expected that only minor modifications
are needed in more accurate detector simulations.
Next, the \ptH\ spectrum is reweighted so as to agree with the spectrum
obtained in Ref.~\cite{Bozzi:2003jy}. A similar procedure is applied to
the main background, the non-resonant WW production (Sect. 4).
Finally, in Sect. 5, the effective experimental \kfac\ is calculated
for the signal and the background and the possible statistical
signal significance is obtained.


\section{$K$-factors and the reweighting technique}
\label{defKfacs}

Generally speaking, the number of events for a given integrated luminosity and a particular process, as
computed at NLO, is given by
\begin{equation}
\label{kincl}
N/\mathcal{L} = \sigma_\mr{NLO} \,\mbox{$=$}\, K_\mr{I} \,\sigma_\mr{LO}
\end{equation}
where $K_\mr{I}$ is called the inclusive \kfac,
which is thus defined as the ratio of the inclusive (total) NLO 
and LO cross sections. At LO, the produced particle (system) X (we have in mind X = H, WW, $\dots$)
has no transverse momentum, while in (N)NLO additional jets lead to a non-vanishing \pt\ spectrum. 
Suppose that a jet veto is applied.
Depending on the jet-detection capabilities 
of the hypothetical experiment one can define an effective parton-level $K$-factor
from the ratio of accepted events at (N)NLO and at LO (for this example all LO events would be accepted).
This effective theoretical (parton-level) $K$-factor is in general smaller than the inclusive $K$-factor \cite{Catani:2001cr}.
   
The definition of effective experimental $K$-factors is different and somewhat more complicated, 
because most experimental simulations are based on LO cross sections, 
supplemented with parton showering through the standard MC programs.
Thanks to additional (mostly soft) jets, an approximate \pt\ spectrum 
of the particular final state is generated.
However, this spectrum is at best an approximation, since only soft and collinear radiation
from the primary parton subprocess can be
generated correctly\footnote{QCD radiation at large transverse momenta
is strongly suppressed and can be accounted for either with matrix-element corrections \cite{Seymour:1994df,Miu:1998ju} 
or by matching the full NLO calculation to the parton shower \cite{mcnlo}.}.

On the other hand, if the (N)NLO calculation has been performed differentially
as a function of a kinematical variable $\xi$,
the above equation can then be rewritten as
\begin{equation}
    N/\mathcal{L} = \int  \frac{d\sigma_\mr{NLO}}{d\xi} d\xi \,\mbox{$=$}
    \,\int K(\xi)  \frac{d\sigma_\mr{MC-LO}}{d\xi} d\xi\, ,
\end{equation}
where the integral goes over the complete possible range of $\xi$, and the $\xi$-dependent \kfac\ is defined as
\begin{equation}
\label{kfactor}
      K(\xi) = \left(\frac{d\sigma_\mr{NLO}(\xi)}{d\xi}    \right) \Big/  
                       \left( \frac{d\sigma_\mr{MC-LO}(\xi)}{d\xi} \right) \; .
\end{equation}

In our case PYTHIA is employed as a leading-order MC program (MC-LO).
It is now possible to study the effects of selection cuts that depend on $\xi$.
In this paper we will be concerned with the case $\xi=\pt$, \pt\ being
the transverse momentum of the Higgs boson (signal) or of the WW pair (background).
The efficiency can be calculated as a function of \pt\ using the 
PYTHIA program. After all cuts are applied we
can also define an inclusive average efficiency to detect process X.
The total number of accepted events within PYTHIA concides
with the sum of the events
that were accepted differentially over the
entire \ptH\ spectrum.

However, if the \pt-dependent $K$-factor defined in
Eq.~(\ref{kfactor}) is applied to correct (reweight) the spectrum,
the total number of accepted events can only be obtained 
from the sum of the differentially accepted weighted events. 
This new number of accepted events at (N)NLO can now be compared 
with the one accepted in the unweighted PYTHIA simulation
and their ratio defines
the effective experimental \kfac\ $K_{\mr{eff}}$.  


\section {Higgs signal selection}
  \label{selection}

The \PY\ MC program is used for the simulation of 
the signal and the different types of relevant backgrounds.
The strategy to separate signal events of the 
type $\mr{pp}\ra \mr{H} \ra \mr{W}^{+}\mr{W}^{-} \ra \ell^{+} \nu \ell^{'-} \bar{\nu}$
from the various backgrounds is based on the ideas described in 
Ref.\ \cite{herbidittmar}.
The signal selection proceeds in two steps.

First, events that contain exactly two isolated and oppositely charged high-\pt\ leptons
(electrons or muons) are selected. These leptons originate mainly from the decays of W bosons.
They should not be back-to-back in the plane transverse to the beam
and their invariant mass should be considerably
 smaller than the Z mass. Furthermore, a substantial  
missing transverse momentum is required.
Essentially, these criteria select only events that contain a pair of W's,
these being either 
signal events or backgrounds from 
non-resonant WW production $\mr{q}\bar{\mr{q}} \ra \mr{WW} \ra \ell^{+} \nu \ell^{'-} \bar{\nu}$, 
from $\mr{t}\bar{\mr{t}} \ra \mr{WbWb} \ra \ell^{+} \nu \ell^{'-} \bar{\nu} \mr{bb}$ 
and $\mr{Wtb} \ra \mr{WWb} \ra \ell^{+} \nu \ell^{'-} \bar{\nu} \mr{b}$.

Following this preselection, the criteria for
the second step further separate  
the Higgs signal events from backgrounds using:
(1) the somewhat shorter rapidity plateau for signal events,
(2) the jet activity in signal events reduced with respect to the 
background from $\mr{t}\bar{\mr{t}}$ production,
(3) the effects of spin correlations and the mass of the resonant and non-resonant 
WW system, resulting in a small opening angle for the lepton--lepton system and a somewhat mass-dependent 
characteristic \pt\ spectrum of the charged leptons.

In detail the following cuts are applied:
\begin{enumerate}
\item The event should contain two leptons, electrons or muons, with opposite charge, 
each with a minimal \pt\ of 20 \gevc\ and a 
pseudorapidity $|\eta|$ smaller than 2.
\item
In order to have isolated leptons, it is required that the 
transverse energy sum from detectable particles (defined as ``stable'' charged
 or neutral particles with a \pt\ larger than 1 \gevc),
found inside a cone of $\Delta R = \sqrt{\Delta \eta^2+\Delta \phi^2)}< 0.5$
around the lepton direction, be smaller than 10\% of the
lepton energy.
Furthermore
the invariant mass of all particles within the 
cone should be smaller than 2 \gevcc,
and at most one additional detectable particle inside a cone of $\Delta R < 0.15$ is allowed.
\item
The dilepton mass, $m_{\ell \ell}$, has to be 
smaller than 80 \gevcc. 
\item
The missing \pt\ of the event, required to balance the 
\pt\ vector sum of the two leptons, should be larger than 20 \gevc.
\item
The two leptons should not be back-to-back in the plane 
transverse to the beam direction. The opening angle 
between the two leptons in  this plane is required to be smaller than 135$^{\circ}$. 
\end{enumerate}

Dilepton events, originating from the decays of W and Z bosons, 
are selected with criteria 1 and 2. Lepton pairs
that originate from the inclusive production of $\mr{Z} \rightarrow \ell \ell (\gamma)$, 
including Z decays to $\tau$ leptons, are mostly removed with criteria 3--5.
 
Starting with this initial set of requirements, the following criteria exploit the 
differences between Higgs events 
and the so-called ``irreducible" background from continuum production 
of $\mr{pp} \ra \mr{W}^{+} \mr{W}^{-}$X events.
\begin{enumerate}
\setcounter{enumi}{5}
\item The opening angle $\phi$
between the two charged leptons in the plane transverse
to the beam should be smaller 
than 45$^{\circ}$
and the invariant mass of the lepton pair should be smaller than 35 \gevcc\footnote{A minimal angle (or mass) of 10$^{\circ}$ (10 GeV) might be 
needed in order to reject badly measured $\Upsilon \rightarrow \mr{e}^{+}\mr{e}^{-} (\mu^{+}\mu^{-})$ decays.
Such a cut would not change the signal efficiency in any significant way.}.
\item
For jets, which are formed with a cone algorithm,
a minimum transverse momentum of 20 \gevc\ is required.
Events with a jet of \ptJ\ larger than
a chosen value $p^{\mr{jet}}_{\mr{Tmin}}$ 
and a pseudorapidity $|\eta^\mr{jet}|$ of less than 4.5 are removed.

\item
Finally, the \pt\ spectrum of the two charged leptons is exploited. For this, the two leptons
are classified according to their \pt\ into \ptmin\ and \ptmax. 
It is found that the \ptmax\ and \ptmin\ distributions
show a Jacobian-peak like structure for the signal, 
which also depends on the simulated Higgs mass. 
In the case of a Higgs mass close to 165 \gevcc,  
\ptmax\ should be between 35 and 50 \gevc, while
the \ptmin\ should be larger than 25 \gevc.
For $M_{\rm H}=140$ \gevc\ the \pt\ of the leptons should be larger than 20 \gevc,
whereas for $M_{\rm H}=180$ \gevc,
the lepton  \ptmax\ has to be larger than 45 \gevc\ and \ptmin\ larger than 25 \gevc.

\end{enumerate}

\begin{figure}[htb]
\begin{center}
\rotatebox{0}{
\epsfig{file=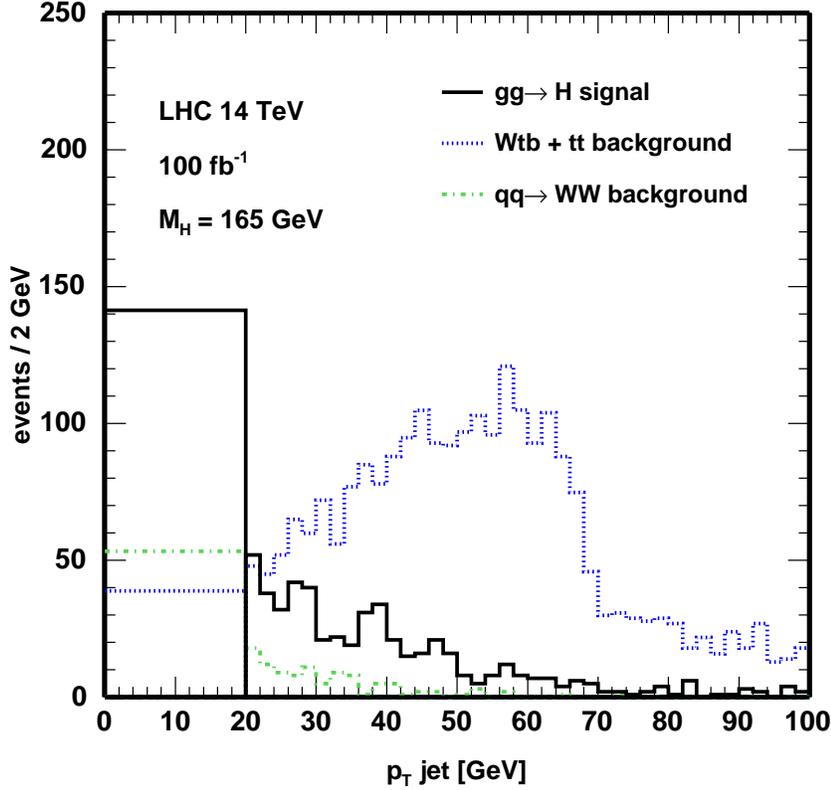,
height=12 cm}}
\end{center}
\caption{Number of accepted signal and background events as a function
of the \ptJ\ of the leading jet. The simulated Higgs
mass is 165 \gevcc. All cuts except the jet veto are applied.   
Events without a reconstructed jet are evenly distributed over the first 10 bins,
since only those jets are counted which have a  reconstructed \ptJ\  larger than 20 \gevc. 
\label{LeadJetpt}
}
\end{figure}

Figure \ref{LeadJetpt} shows the \ptJ\ of the 
hardest (leading) reconstructed jet for signal and background events from \PY\,
after applying the cuts 1 to 6 and cut 8.
The events with no reconstructed jet are equally distributed over the 10 bins between 0 and
20 \gevc.
As can be seen from Fig.\ \ref{LeadJetpt}, the particular choice of the
cut value for the jet transverse momentum 
does not seem to be very critical for the observation of a Higgs signal,
but it needs to be studied in detail if 
a precision cross section measurement is 
envisaged, or if the search is to be extended to much lower values of the Higgs mass 
with smaller signal-to-background ratios.

{\small
\begin{table}[ht]
\begin{center}
\begin{tabular}{|c|c|c|c|c|}
\hline
   & & \multicolumn{3}{c|}{Number of events, $\mathcal{L}=5\, \mr{fb}^{-1}$} \\[0.2cm]
\cline{3-5}
& $\sigma_{\mr{LO\  PYTHIA}}\times \mr{BR}^2$ &All cuts except& & \\
  Process &  [pb] & cuts on
$\pt\ \mr{lepton}$&  35$<\ptmax<$50& 25$<\ptmin$ \\[0.2cm]
\hline
$\mr{gg} \rightarrow \mr{H} \rightarrow \mr{WW}$ &1.06&176  & 110& 80\\

\hline
$\mr{qq} \rightarrow \mr{WW}$&7.38&243  & 83& 30\\
$\mr{t}\bar{\mr{t}}$ &52&47  & 15& 5\\
 Wtb &5.2& 87 &46 &26 \\
\hline
\end{tabular}\vspace{0.3cm}
\caption{\label{165}Cross sections obtained with PYTHIA  for signal and backgrounds, and expected number
of events after applying various cuts, for $M_{\rm H}=165$ GeV and an integrated luminosity of 5 fb$^{-1}$.
In all cases the leptonic branching ratios of both W bosons, $\mr{W} \rightarrow\ell \nu$ with $\ell = \mr{e}, \mu, \tau$ are included.
}

\end{center}
\end{table}
}

Applying all selection criteria, including the optimized lepton \pt\ cuts, 
an accepted 
cross section of 15.9 fb for the Higgs signal with a mass of 165 GeV can be expected,
above a background of 12.3 fb.
In principle, to estimate the complete signal rate,
the contribution from vector boson fusion,
which, for $M_{\rm H}=165$ GeV, is about $0.7$ fb, should be included.
For the purpose of this paper, this contribution will be neglected in the following.

The LO cross sections indicate that for Higgs masses near 165 GeV
a statistically significant signal should 
be observable with a luminosity of slightly more than 1 fb$^{-1}$.
The relevant PYTHIA cross sections for the signal and backgrounds as well as some event rates 
expected for a luminosity of 5 fb$^{-1}$ are given in Table \ref{165}.

\begin{figure}[htb]
\begin{center}
\rotatebox{0}{
\epsfig{file=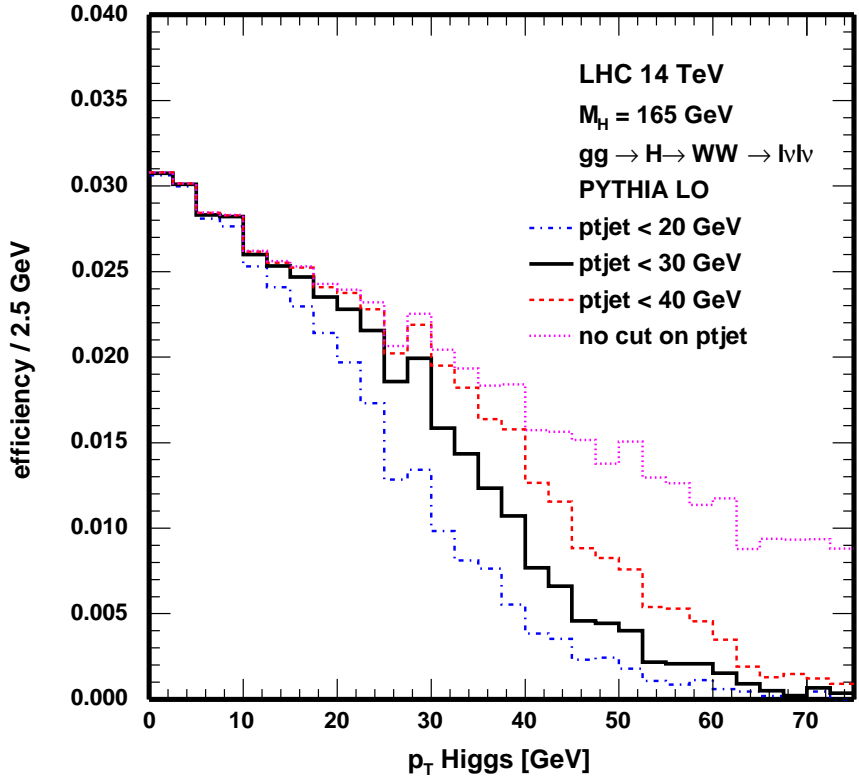,
height=12 cm}}
\end{center}
\caption{Signal selection efficiency as a function of the Higgs transverse momentum, 
for  a Higgs mass of 165 \gevcc\ and three different jet veto cuts. For completeness, 
the efficiency curve for all cuts, excluding the jet veto, is also shown.
\label{efficiency}}
\end{figure}

As already mentioned in the introduction, the
analysis mainly selects signal events 
with low \ptH.
The efficiency to detect a Higgs boson
with these selection criteria, 
defined as the ratio of all accepted 
over all generated events, can be studied 
as a function of the generated transverse momentum of the Higgs. 
The results for different jet veto cuts ($p_\mr{Tmin}^\mr{jet}$ = 20, 30 and 40 \gevc)
are shown in Fig.\ \ref{efficiency}. 
As expected, Higgs events with large \ptH\ are
almost always rejected
with the proposed criteria,
and the efficiency drops quite quickly as \ptH\ reaches
the value of the jet veto  $p_\mr{Tmin}^\mr{jet}$.

\section{Signal and background \pt\ spectra}

At the lowest order in QCD perturbation theory, the Higgs boson
is produced with vanishing transverse momentum. Thus, in order to generate a non-vanishing $\ptH$,
NLO corrections should be considered. At transverse momenta $\ptH$
of the order of $M_{\rm{H}}$ 
the perturbative expansion is reliable,
being controlled by a small expansion parameter $\as(M_{\mr{H}}^2)$.
NNLO corrections have been computed,
in the large-$m_{\mr{top}}$ 
approximation,
first numerically \cite{deFlorian:1999zd}
and later analytically \cite{Ravindran:2002dc,Glosser:2002gm}
\footnote{Contrary to the convention adopted
in Refs.~\cite{deFlorian:1999zd,Ravindran:2002dc,Glosser:2002gm},
we use here the classification of perturbative orders
based on the total cross section: the $p_{\rm T}$ spectrum starts at NLO.}.

When $\ptH\ $ is much smaller than $M_{\mr{H}}$ the
convergence of the fixed-order expansion is spoiled, as
the coefficients of the perturbative series in $\as(M_{\mr{H}}^2)$ are enhanced
by powers of large logarithmic terms, $\ln^m (M_{\mr{H}}/\ptH)$.
These terms must be resummed
to all orders to give a perturbative prediction valid down to small \ptH.

In Ref.~\cite{Bozzi:2003jy} the resummation of these logarithmic contributions
is performed up to next-to-next-to-leading logarithmic accuracy (NNLL)
and matched to the fixed-order (NNLO) result valid at large \ptH, in order to avoid double counting.
We thus obtain a prediction that is always as good as the fixed-order prediction, but much better in the
small-$\ptH$ region.
Note that the resummation of logarithmically enhanced contributions
is approximately performed by standard parton shower MC programs, which should thus account
for the shape of the distribution in the small-$\ptH$ region.

In the following we will use results obtained with the numerical program
of Ref.~\cite{Bozzi:2003jy} at NNLL+NNLO accuracy. 
The formalism of Ref.~\cite{Bozzi:2003jy} implements a unitarity constraint,
such that the integral of the distribution is the total NNLO cross section \cite{higgsnnlo}.
At variance with the calculation of Ref.~\cite{Bozzi:2003jy},
the \ptH\ spectrum is here obtained
using the MRST2002 NNLO \cite{Martin:2003es} parton distributions and
$\as$ computed in the three-loop approximation.

\begin{figure}[htb]
\begin{center}
\rotatebox{0}{
\epsfig{file=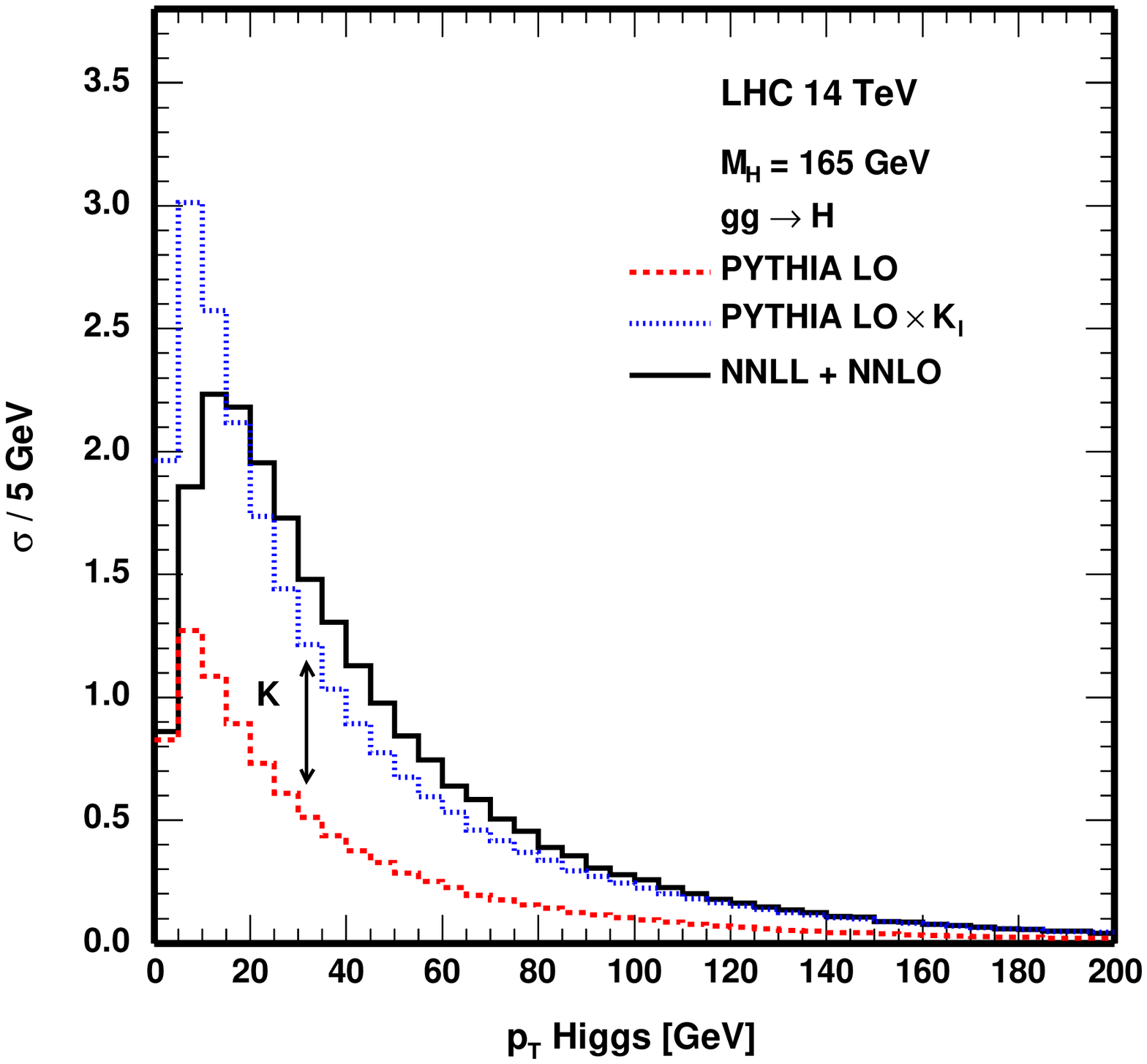,
height=12 cm}}
\end{center}
\caption{The Higgs production cross section for 
$\rm{gg}\ra\rm{H}$, as a function of the 
Higgs transverse momentum \ptH, for a Higgs mass of 165 \gevcc,
  obtained with \PY\ and with the NNLL+NNLO calculation.
The spectrum from \PY\ rescaled with the inclusive $K$-factor
is also shown for comparison.
\label{ptHspectra}
}
\end{figure}

The expected Higgs \ptH\ spectra for $M_{\mr{H}}=165$ GeV from
\PY\ and from the resummed calculation are shown 
in Fig.\ \ref{ptHspectra}. It can be seen that PYTHIA provides a softer \ptH\ spectrum and 
differs from the perturbative
calculation over the whole range of $\ptH$
The ratio between the two spectra can be used to 
define the \ptH-dependent \kfac, $K(\ptH)$ (see Section \ref{defKfacs}).

\begin{figure}[htb]
\begin{center}
\rotatebox{0}{
\epsfig{file=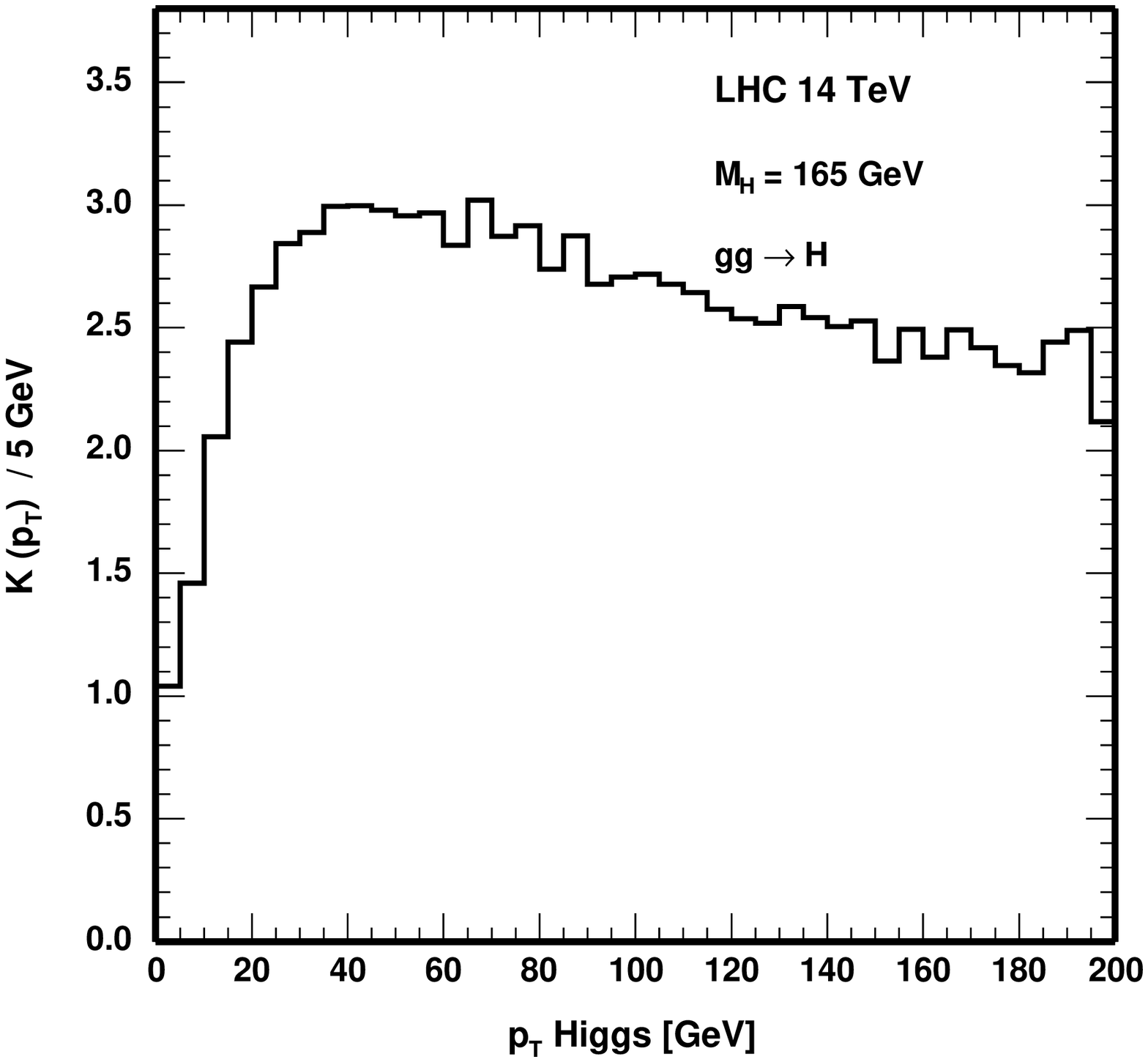,
height=12 cm}}
\end{center}
\caption{The \ptH\ dependence of the \kfac, as defined in  Section \ref{defKfacs}.
\label{Kptdep}
}
\end{figure}

This \ptH-dependent \kfac\ rises from approximately 1 at small \ptH\,
to 3 at a \ptH\ around 50 \gevc, and then decreases again to about 2.2 at a \ptH\ of 200 GeV,   
as shown in Fig.\ \ref{Kptdep}.
Note that at relatively large transverse momenta,
the \PY\ event generator is supplemented
with hard matrix-element corrections \cite{Miu:1998ju},
thus explaining the approximately flat $K$-factor at large \ptH, but
the normalization is still fixed to LO.

The \ptH-dependent \kfac\ can be used to apply a weight to events generated
with PYTHIA.
The idea of the reweighting procedure is based on the assumption 
that the kinematics of Higgs events for a particular \ptH\ 
is reasonably well described by \PY\
and that the efficiency of the cuts is computed correctly.
Since the \pt\ spectrum is generated by multiple radiation from
the incoming partons, the rapid variation
of the \kfac\ for $\ptH\ltap$ 40 GeV in Fig.\ \ref{Kptdep}
could suggest an improper treatment of
the effect of a jet veto in PYTHIA.
In order to check the reliability of
our reweighting procedure we have compared the efficiency
of a jet veto
with the one obtained with HERWIG \cite{hw},
which is known to provide a better description of
the \ptH\ spectrum in the small-\ptH\ region \cite{Dobbs:2004bu}.
When $\ptH\ltap 40$ GeV,
the efficiencies differ by less than $5 \%$,
thus confirming the validity of our approach.


Consequently, it is possible to obtain an approximation
for the (N)NLO distributions 
of the kinematic observables used to select the final state by simply
reweighting each PYTHIA event in such a way that
the new \ptH\ spectrum matches the one from the QCD calculation.
Of course, the \ptH-dependent signal efficiency is not altered by the 
reweighting.

A similar procedure is applied for the main background, the continuum production of WW pairs.
Here transverse momentum spectra obtained with \PY\ are reweighted according to
QCD predictions at next-to-leading-logarithmic (NLL) accuracy,
which are matched to the perturbative NLO result
\cite{Campbell:1999ah,Dixon:1999di},
valid at large transverse momenta $\ptWW$ of the WW pair.
For this calculation\footnote{The NLL resummed WW cross section is
computed according to the formalism of Refs.~\cite{Catani:2000vq,Bozzi:2003jy}.
The NLO result used for the matching is obtained
with the MCFM package \cite{Campbell:1999ah}.
More details on these results will be given elsewhere.}
MRST2002 NLO densities and a running $\as$ in
the two-loop approximation are used, so that
the integral of the spectrum is fixed to the total NLO cross section \cite{wwnlo}.

\begin{figure}[htb]
\begin{center}
{\epsfig{file=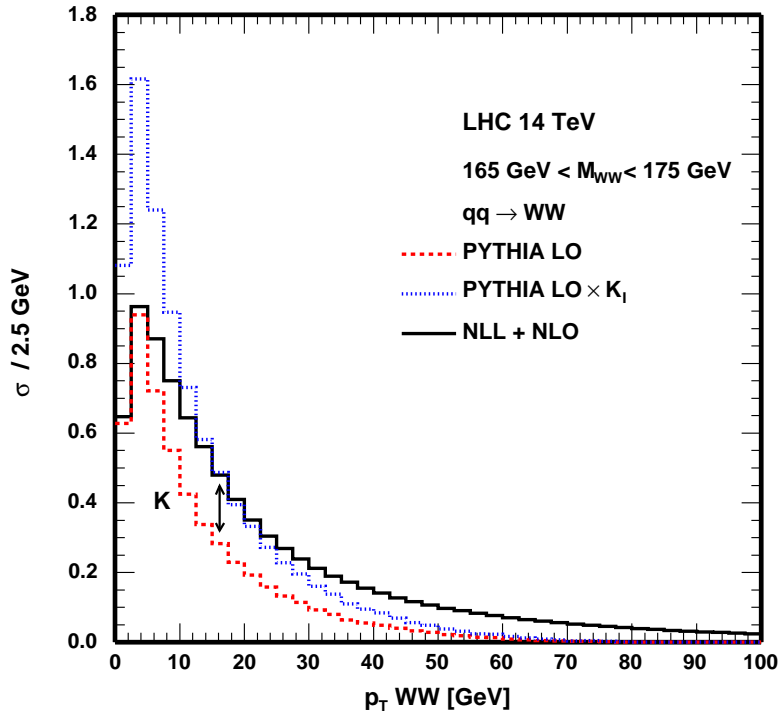,
height=11 cm}}
\end{center}
\caption{The \pt\ spectrum of the non-resonant WW system
with a mass of $170\pm 5$ GeV, obtained from PYTHIA 
and from the NLL+NLO calculation.
The spectrum from \PY\ rescaled with the inclusive $K$-factor
is also shown for comparison.
\label{ptwwspectra}
}
\end{figure}

\begin{figure}[htb]
\begin{center}
\rotatebox{0}{
\epsfig{file=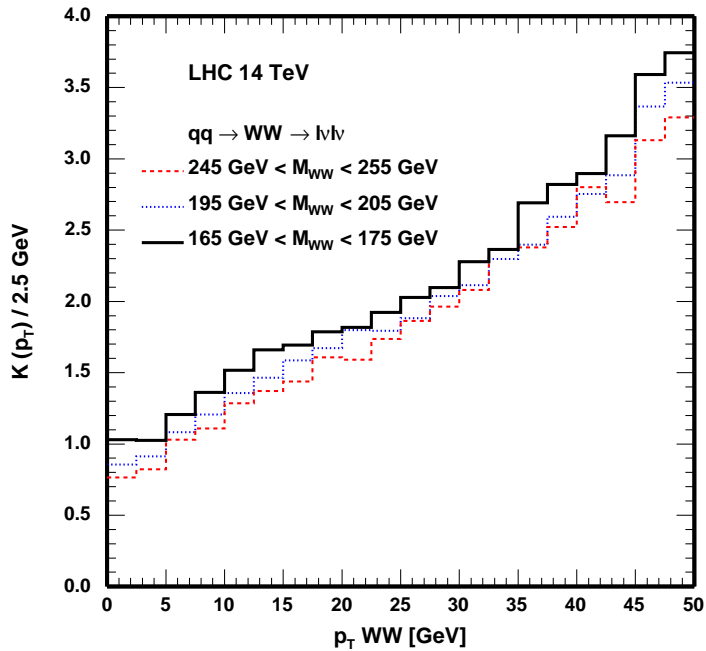,
height=10 cm}}
\end{center}
\caption{ The \pt\ dependence of the \kfac\ for the non-resonant WW system and three 
different WW mass intervals.  
\label{ptwwratios}
}
\end{figure}

In order to compare the \pt-dependent WW spectrum from PYTHIA
with the one from the higher-order calculation,
the dependence on the mass of the WW system  has to be taken into account. 
This is done for three different mass intervals,
$170\pm 5$ GeV, $200\pm 5$ GeV and $250\pm 5$ GeV,
which cover the mass range where the WW events  
are a potential background for the Higgs signal,
using the selection criteria described above.
The expected \ptWW\ spectra from the two calculations in the WW mass
range of $170\pm 5$ GeV
are shown in Fig.\ \ref{ptwwspectra}.
The corresponding
$K$-factors, as a function of $\ptWW$
for the three different WW mass intervals
are shown in Fig.\ \ref{ptwwratios}.

The difference between the \ptWW\ spectrum in PYTHIA and the one
calculated in NLL+NLO QCD 
are particularly large for large transverse momenta.
This is because, contrary to the Higgs signal,
for WW production no hard matrix-element corrections are
applied in \PY, and thus the corresponding spectrum falls rather quickly as
$\ptWW$ increases.

For the analysis described in this paper
only the events with relatively small \pt\ are relevant and the corresponding event weights for the 
non-resonant WW production 
are found to increase from about 1 at small transverse momentum
to almost 4 at a transverse momentum of 50 GeV,
slightly depending on the mass of the WW system.
However, since most of the relevant WW continuum background comes from
events with an invariant mass around threshold and 
relatively low transverse momentum, 
we take as an approximate weighting factor for the WW events
the one obtained for the
mass range $170\pm 5$ GeV.
As can be seen from Fig.\ \ref{ptwwratios}, this will slightly overestimate the WW background.

For the $\mr{t}\bar{\mr{t}}$ and Wtb background, the \pt\ spectrum is taken from PYTHIA, 
while the corresponding cross sections are both simply scaled by a constant \kfac\ of 1.5, so that the 
rescaled $\mr{t}\bar{\mr{t}}$ cross section from \PY\ matches the inclusive NLO $\mr{t}\bar{\mr{t}}$ 
cross section \cite{Nason:1987xz,sigmanlott}.
Although this is certainly a crude approximation, it is applied only to the
less important $\mr{t}\bar{\mr{t}}$ and Wtb backgrounds. Moreover, as we will
discuss in the next section, the accuracy with which the background is known
is less important with respect to the one of the signal,
as far as the statistical significance is concerned.

\section{Results}
 \label{results}

\begin{figure}[htb]
\begin{center}
\rotatebox{0}{\epsfig{file=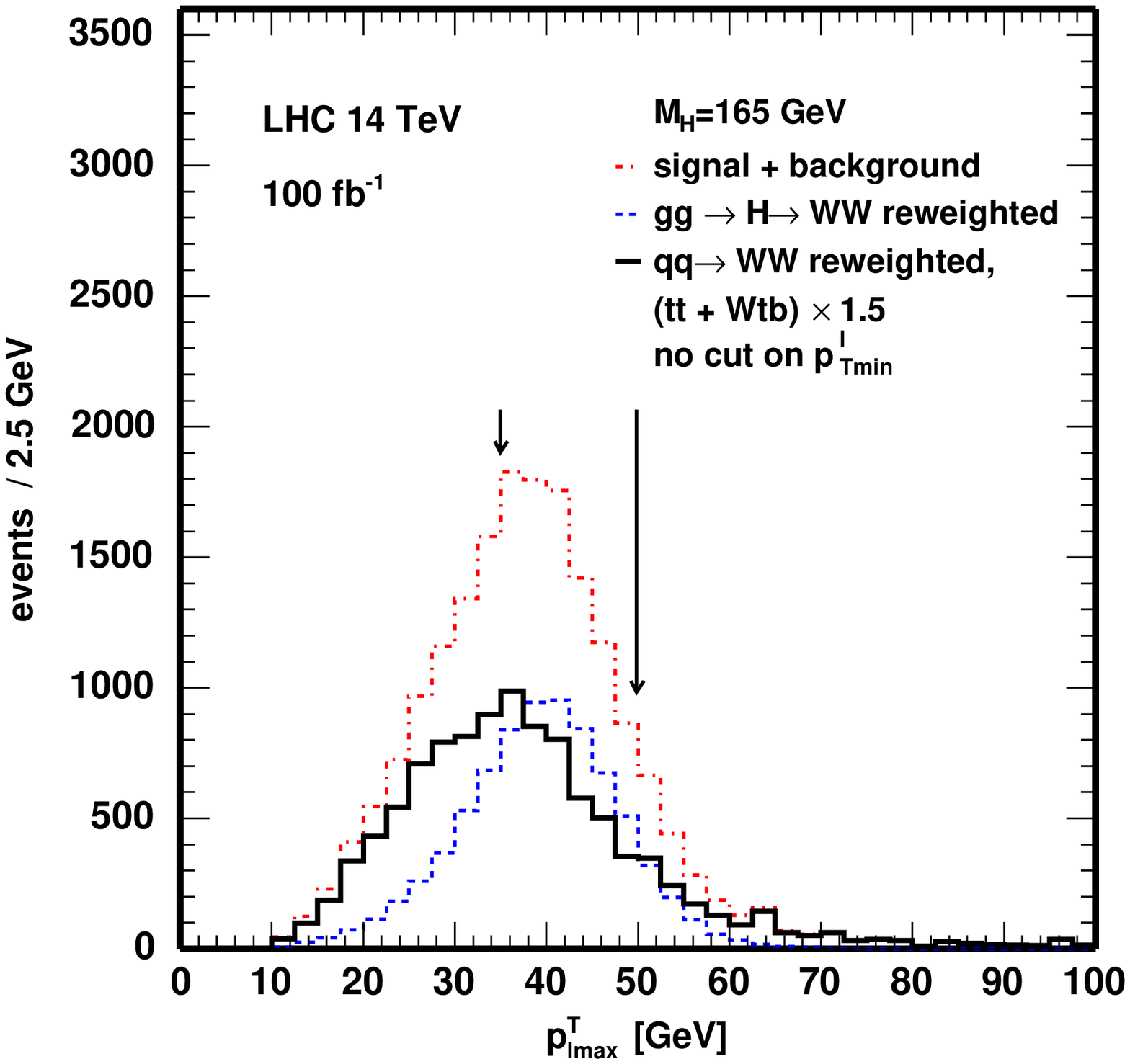,width=9. cm} 
\epsfig{file=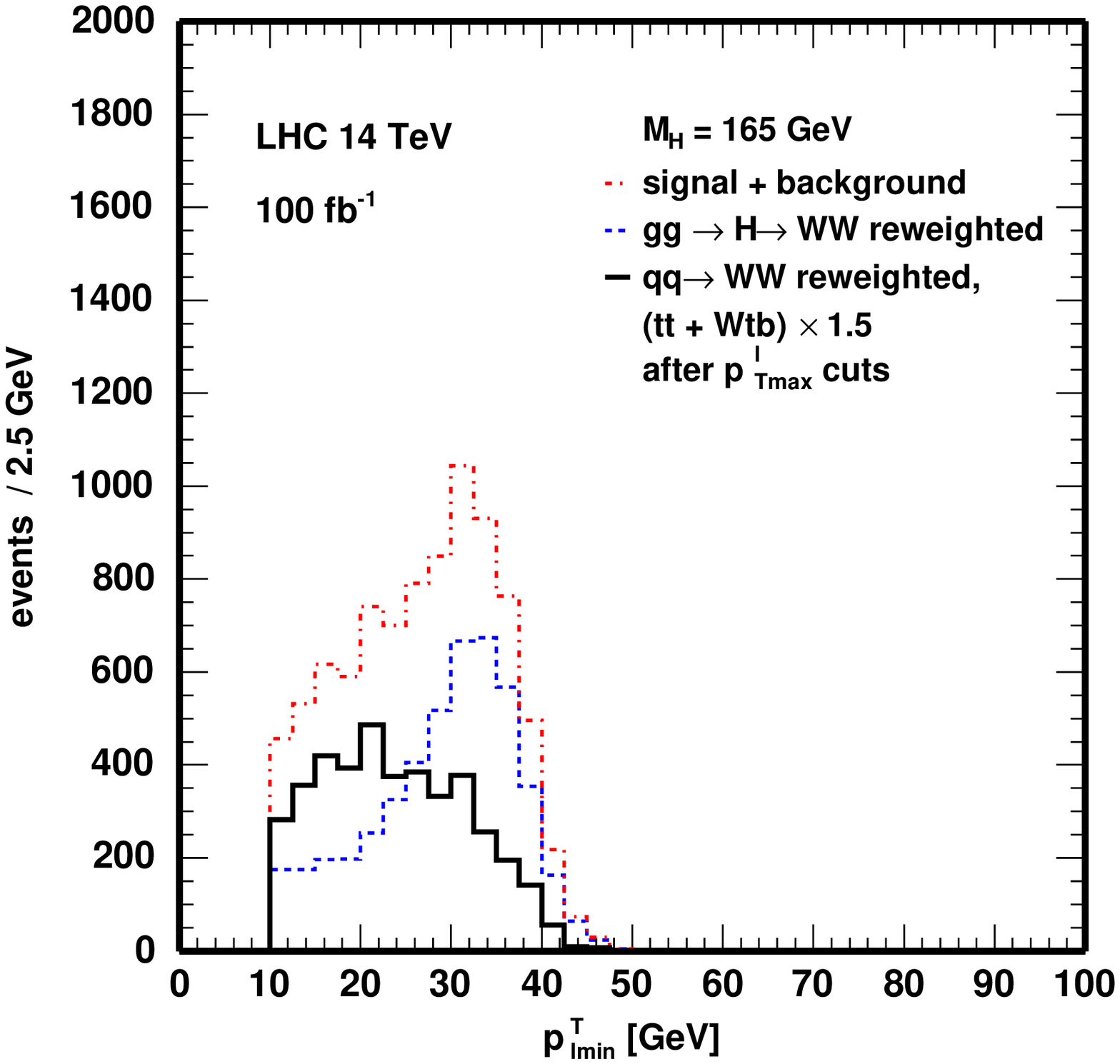,width=9. cm}}
\end{center}
\caption{Transverse momentum spectra of the leading lepton (left) and the
lepton with the smaller \pt\ (right) from
$\rm{gg}\ra\rm{H}\ra\rm{WW}\ra \ell^{+} \nu \ell^{'-} \bar{\nu}$
and from the considered backgrounds,
as obtained from \PY\ with event reweighting.
The expected background
from non-resonant W-pair production is reweighted using
the \pt-dependent weighting factor, while the 
ones from ${\rm t}{\bar {\rm t}}$ and Wtb
are simply scaled by a factor of 1.5.
The \ptmin\ spectrum in the right plot is obtained after all cuts,
including the cut on \ptmax, as indicated by the arrows in the left plot.}
\label{leptonspectra}
\end{figure}

The effective experimental \kfac\ can be computed from the sum of the accepted 
cross section over all the \pt\ bins.
The numbers for a Higgs mass of 165 \gevcc\ 
are given in Table \ref{table165}, both for
the NNLL+NNLO and for the \PY\ prediction. The signal efficiency after the selection described in
Section \ref{selection} is also given in Table \ref{table165}.
The efficiency vanishes for \ptH\
above 65 \gevc. Therefore, although the \kfac\ for the bin $65 < \ptH < 70$ \gevc\ is about 3, it will not 
contribute when computing the effective experimental \kfac\  $K_{\mr{eff}}$.

From the integration over all \ptH\ bins, the inclusive $K$-factor
with respect to \PY,
without any selection cuts,
is found to be $K_\mr{I} = 2.37$.
This is roughly 15\% larger than
$K_\mr{eff}=2.04$, which is obtained after all cuts are applied, 
including the jet veto of 30 \gevcc. 
This means that the number of accepted reweighted events is a factor of 2.04 larger than in the unweighted case. 
Similar numbers are obtained for other Higgs masses. The results for Higgs masses of 140 GeV and 180 GeV 
are given in Table \ref{table140}.
The estimated effective \kfac\ for the WW background, integrating over the entire  
WW mass spectrum and using the \pt-dependent weighting factor determined for the WW mass interval of 
165--175 GeV, is found to be 1.36. 
Considering only the WW mass interval from 165--175 GeV, the effective \kfac\ would be 1.44, which is
about 18\% lower than the inclusive \kfac\
for this WW mass interval.

{\small
\begin{table}[ht]
\begin{center}
\begin{tabular}{|c|cccc|}
\hline
&   & &   &\\
&  &$\!\!\!\!M_{\rm H}$ = 165 GeV& &  \\
&   & &   &\\
\ptH\,[GeV]&  $\sigma_{\mr{NNLO+NNLL}}$\,[pb]& $\sigma
_{\mr{PYTHIA}}$\,[pb]&$K$&$\epsilon$\,[\%]  \\
&   & &  &\\
\hline
0--5   & 0.861& 0.828 &1.040&3.0 \\
5--10  & 1.856& 1.272&1.460&2.8\\
10--15 & 2.233& 1.086&2.057&2.6\\
15--20 & 2.180& 0.892&2.443&2.4\\
20--25 & 1.954& 0.733&2.667&2.2\\
25--30 & 1.729& 0.608&2.842&1.9\\
30--35 & 1.481& 0.513&2.889&1.5\\
35--40 & 1.306& 0.436&2.995&1.2\\
40--45 & 1.129& 0.377&2.997&0.7\\
45--50 & 0.976& 0.327&2.980&0.4\\
50--55 & 0.843& 0.285&2.958&0.3\\
55--60 & 0.746& 0.251&2.968&0.2\\
60--65 & 0.637& 0.225&2.836&0.1\\
65--70 & 0.585& 0.194&3.020&0.0\\
70--80 & 0.960& 0.332&2.892&0.0\\
80--90 & 0.744& 0.265&2.808&0.0\\
90--100& 0.584& 0.217&2.691&0.0\\
100--200&2.276& 0.809&2.813&0.0\\
\hline
&   & &   &\\
Total & 23.08 & 9.74& & \\
&   & &   &\\
\hline
&   & &   &\\
 & {$K _\mr{I} = 2.37$} &  {$K_{\rm eff}$ = 2.04}  & &\\
&   & &   &\\
\hline
\end{tabular}
\vspace{0.3cm}
\caption{\label{table165}Higgs production cross sections as a function of the
Higgs transverse momentum \ptH,
  obtained in NNLL+NNLO QCD and with \PY, for a Higgs mass of 165 \gevcc.
The third and fourth columns list the \pt-dependent $K$-factor and the signal
selection efficiency, respectively.}
\end{center}
\end{table}
}

{\small
\begin{table}[ht]
\begin{center}
\begin{tabular}{|c|ccc|ccc|}

\hline
&   & &  & & & \\
&  &\hskip -1.5cm $M_{\mr{H}}$ = 140 GeV & & &\hskip -1.5cm $M_{\mr{H}}$ = 180 GeV & \\
&   & &  & & & \\
\ptH\,[GeV] & $\sigma_{\mr{NNLO+NNLL}}$\,[pb]& $\sigma_{\mr{PYTHIA}}$\,[pb]& $\epsilon$\,[$\%$]&
$\sigma _{\mr{NNLO + NNLL}}$\,[pb]& $\sigma _{\rm{PYTHIA}}$\,[pb] & $\epsilon$\,[$\%$]\\
&   & &  & & & \\
\hline
0--5   & 1.342  & 1.230 &2.2  &0.680 &0.674 &0.7\\
5--10  & 2.727  & 1.805&2.1  &1.420 &1.052&0.8\\
10--15 & 3.430  & 1.495&2.1  & 1.809&0.918&0.8\\
15--20 & 3.162  &1.208 &2.1  &1.762&0.759 &0.8\\
20--25 & 2.816  &0.979 &1.9  &1.642&0.633 &0.8\\
25--30 & 2.436  &0.800 &1.7  &1.421& 0.526&0.8\\
30--35 & 2.094  &0.668 &1.6  &1.250&0.440 &0.8\\
35--40 & 1.781  & 0.567&1.3  &1.081 &0.379 &0.6\\
40--45 & 1.370  & 0.483&1.0  &0.974&0.330&0.5 \\
45--50 & 1.445  & 0.411&0.7  &0.820&0.287&0.3 \\
50--60 & 2.088  & 0.675& 0.5&1.367&0.473&0.2\\
60--70 & 1.578  & 0.516&0.2 &1.067 & 0.369&0.1 \\
70--80 & 1.198  & 0.407&0.1 & 0.829& 0.297&0.1\\
80--90 & 0.934  & 0.316&0.0 &0.664 &0.239 &0.0\\
90--100& 0.724  & 0.257&0.0  &0.530&0.194 &0.0\\
100--200& 2.665  & 1.003&0.0 &2.064& 0.830&0.0\\
\hline
&   & &  & & & \\
Total & 31.79 & 12.82 &  &19.38  & 8.40 &  \\
&   & &  & & & \\
&  K$_\mr{I}$=2.48 & K$_{\rm eff}$= 2.25 & & K$_\mr{I}$=2.30 & K$_{\rm eff}$=2.03 &
\\
&   & &  & & & \\
\hline
\end{tabular}\vspace{0.3cm}
\caption{\label{table140}Higgs production cross sections as a function of the
Higgs transverse momentum \ptH,
obtained in NNLL+NNLO QCD and with \PY, for a Higgs mass of 140 \gevcc\
(left) and 180 \gevcc\ (right).
The signal selection efficiency is also given.}
\end{center}
\end{table}
}


\begin{table}[ht]
\begin{center}
\begin{tabular}{|c|c|c|c|c|c|c|}
\hline
& & && & &\\
$M_{\rm H}$ [GeV] & $S$ & WW & Wtb & ${\rm t}{\bar {\rm t}}$ &$S/B$& $S/\sqrt{B}$\\
& & & && &\\
\hline
140 & 106 & 158 & 87 & 34 & 0.38& 6\\
\hline
165 & 162 & 44 & 40  & 7 &1.78& 17\\
\hline
180 & 48 & 23 & 17  &  7&1.02& 7\\
\hline
\end{tabular}
\vspace{0.3cm}
\caption{\label{tablenew}Number of signal and background events and corresponding statistical significance for an integrated luminosity of 5 fb$^{-1}$.}
\end{center}
\end{table}

Following this procedure,
the next step consists in 
calculating the luminosity requirements for the observation of a Higgs signal 
with a statistical significance of five standard deviations. 
In order to calculate the potential statistical significance of a signal
it is usually assumed
that the background is accurately known. With this assumption and requiring   
a significance of five standard deviations, the number of signal 
events $S$ has to be equal to
5$\times \sqrt{B}$
\footnote{For small event numbers, this has to be replaced 
with a probability calculation based on Poisson statistics.},
$B$ being the number of background events.
Thus, ignoring systematic uncertainties,
the accurate knowledge of the signal cross section is more important
than the absolute value for the background.

However, once systematic uncertainties are considered, the accurate knowledge of the background 
becomes relevant, especially if $S/B$ is smaller than 1.
In addition,
uncertainties from the luminosity
and the parton distribution functions, as well as from
the experimental efficiency
need to be considered in detail.
Most of these uncertainties can only be determined accurately once the first LHC data are obtained.

The transverse momentum spectra of the two leptons, \ptmax\ and \ptmin, 
after the reweighting and with all cuts applied,
are shown in Fig.\ \ref{leptonspectra} for $M_{\mr{H}}=165$ GeV.

The accepted events for signal and backgrounds for
$M_{\rm H}=140$, $165$, $180$ GeV and
the corresponding statistical significance for 5 fb$^{-1}$
are given in Table \ref{tablenew}. 
We see that
signal-to-background ratios
between about 1:2 and 2:1 can be obtained in the mass range under consideration. 
In the case of $M_{\rm H}=165$ GeV, a statistical significance of five standard deviations
can already be achieved with an integrated luminosity of about 0.4 fb$^{-1}$. 

In order not to 
affect the discovery potential in any significant way, the
systematic uncertainties on the background have to be controlled 
to better than about 10--20\%. 
This may well be achievable knowing
that (1) the shape of various distributions can be calculated with  
good accuracy, and that (2) kinematic regions with small signal contributions can be
isolated and used to normalize the potential backgrounds
with systematic accuracies of perhaps 5--10\%.


\section{Summary}

We have performed a simulation of the SM Higgs boson search at the LHC
in the channel $\mr{gg}\ra \mr{H} \ra \mr{W}\mr{W}\ra l\nu l\nu$.
QCD corrections have been included
by using a reweighting procedure, allowing us to combine
event rates obtained with \PY\ with
the most up-to-date theoretical predictions
for the transverse momentum spectra for the $\mr{gg}\ra \mr{H}$ signal and its main WW background.

The reweighting method has been used to compare the 
experimental sensitivity to find the Higgs
estimated with PYTHIA with the one obtained by taking
higher-order QCD corrections into account.
In particular, the 
effect of a jet veto on the weighted and unweighted
events has been investigated.
Using this procedure and a Higgs mass of 165 \gevcc, the effective experimental \kfac\
is only about
$15\%$ smaller than the inclusive \kfac.
From these results the Higgs discovery potential for the
channel $\mr{pp}\ra \mr{H} \ra \mr{W}^{+}\mr{W}^{-} \ra \ell^{+} \nu \ell^{'-} \bar{\nu}$ 
is found to be significantly increased.
Consequently, signals with a statistical significance of five standard deviations 
should be observable for  
a SM Higgs boson with masses 
between 140 and 180 \gevcc\  after the first few $\mr{fb}^{-1}$ of integrated luminosity.

The reweighting technique proposed in this paper can be applied
to other final states 
and the results should be particularly accurate for hard scattering processes with little additional jet activity.

\vspace{1cm}

\section*{Acknowledgements}

We thank Stefano Catani, Daniel de Florian, Stefano Frixione and Andre Holzner
for useful comments and discussions.
One of us (MG) would like to thank John Campbell
for his help in the use of MCFM.

\end{document}